\documentclass[%
 reprint,
superscriptaddress,
 amsmath,amssymb,
 aps,
 prl,
]{revtex4-1}

\usepackage{graphicx}% Include figure files
\usepackage{dcolumn}% Align table columns on decimal point
\usepackage{bm}% bold math
\usepackage[colorlinks= true, linkcolor=blue, citecolor=blue, urlcolor=blue]{hyperref}% add hypertext capabilities
\usepackage{lipsum}
\usepackage{bbold}
\usepackage{mathtools}
\usepackage[normalem]{ulem}

\def\bea{\begin{eqnarray}}
\def\eea{\end{eqnarray}}

\usepackage{xcolor}

\begin{document}

%\preprint{APS/123-QED}

\title{Towards Stirling engine using an optically confined particle subjected to asymmetric temperature profile}

\author{Gokul Nalupurackal}
\affiliation{Department of Physics, Quantum Centres in Diamond and Emergent Materials (QuCenDiEM)-group, Micro Nano and Bio-Fluidics (MNBF)-Group,
IIT Madras, Chennai 600036, India}

\author{Muruga Lokesh}
\affiliation{Department of Physics, Quantum Centres in Diamond and Emergent Materials (QuCenDiEM)-group, Micro Nano and Bio-Fluidics (MNBF)-Group,
IIT Madras, Chennai 600036, India}

\author{Sarangi Suresh}
\affiliation{Department of Physics, Quantum Centres in Diamond and Emergent Materials (QuCenDiEM)-group, Micro Nano and Bio-Fluidics (MNBF)-Group,
IIT Madras, Chennai 600036, India}

\author{Srestha Roy}
\affiliation{Department of Physics, Quantum Centres in Diamond and Emergent Materials (QuCenDiEM)-group, Micro Nano and Bio-Fluidics (MNBF)-Group,
IIT Madras, Chennai 600036, India}
\author{Snigdhadev Chakraborty}
\affiliation{Department of Physics, Quantum Centres in Diamond and Emergent Materials (QuCenDiEM)-group, Micro Nano and Bio-Fluidics (MNBF)-Group,
IIT Madras, Chennai 600036, India}

 \author{Jayesh Goswami}
\affiliation{Department of Physics, Quantum Centres in Diamond and Emergent Materials (QuCenDiEM)-group, Micro Nano and Bio-Fluidics (MNBF)-Group,
IIT Madras, Chennai 600036, India}

\author{Gunaseelan M.}
\affiliation{Department of Physics, Quantum Centres in Diamond and Emergent Materials (QuCenDiEM)-group, Micro Nano and Bio-Fluidics (MNBF)-Group,
IIT Madras, Chennai 600036, India}
 

\author{ Arnab Pal}
\email{arnabpal@imsc.res.in}
\affiliation{The Institute of Mathematical Sciences, CIT Campus, Taramani, Chennai 600113, India \& 
Homi Bhabha National Institute, Training School Complex, Anushakti Nagar, Mumbai 400094,
India
}

\author{Basudev Roy}%
 \email{basudev@iitm.ac.in}
\affiliation{Department of Physics, Quantum Centres in Diamond and Emergent Materials (QuCenDiEM)-group, Micro Nano and Bio-Fluidics (MNBF)-Group,
IIT Madras, Chennai 600036, India}

\begin{abstract}
The realization of microscopic heat engines has gained a surge of research interest in statistical physics, soft matter, and biological physics. A typical microscopic heat engine employs a colloidal particle trapped in a confining potential, which is modulated in time to mimic the cycle operations. Here, we use a lanthanide-doped upconverting particle (UCP) suspended in a passive aqueous bath, which is highly absorptive at 975 nm and converts NIR photons to visible, as the working substance of the engine. When a single UCP is optically trapped with a 975 nm laser, it behaves like an active particle by executing motion subjected to an asymmetric temperature profile along the direction of propagation of the laser. The strong absorption of 975 nm light by the particle introduces a temperature gradient and results in significant thermophoretic diffusion along the temperature gradient. However, the activity of the particle vanishes when the trapping wavelength is switched to 1064 nm. We carefully regulate the wavelength-dependent activity of the particle to engineer all four cycles of a Stirling engine by using a combination of 1064 nm and 975 nm wavelengths. Since the motion of the particle is stochastic, the work done on the particle due to the stiffness modulation per cycle is random. We provide statistical estimation for this work averaged over 5 cycles which can be extended towards several cycles to make a Stirling engine. Our experiment proposes a robust set-up to systematically harness temperature which is a crucial factor behind building microscopic engines. 
\end{abstract}

\pacs{Valid PACS appear here}% PACS, the Physics and Astronomy
\maketitle
%\onecolumngrid

\onecolumngrid

\section{\label{sec:levelintro}Introduction}

Microscopic heat engines are well-known as prototypical tools for the investigation of energy transduction in fluctuation-dominated regimes. The core studies on this specific area are vividly about the realizations of heat engines that operate between different thermal reservoirs. These engines, however, operate in non-equilibrium conditions such that stochastic thermodynamics provides a useful description in their quantification \cite{intro1, intro2,seifert2012stochastic,jarzynski2011equalities,polettini2015efficiency,verley2014universal,Kumari2020,intro5, intro6,martinez2017colloidal}. This is indeed crucial since significant research has been done on artificial \cite{browne2006making, balzani2000artificial} and biological microscopic engines and motors \cite{howard2002mechanics,ramaiya2017kinesin} with exciting applications. Recent studies include working principles and operational strategies of many such microscopic non-equilibrium systems namely the Brownian Stirling engine \cite{intro1,clemensnature}, active Stirling engine \cite{krishnamurthy2016micrometre,holubec2020active}, Brownian Carnot's engine \cite{martinez2016brownian,frim2022optimal} and information engine \cite{admon2018experimental}. Generically, a prototypical set-up of such studies includes a micro-particle (which takes the role of the working substance) confined in a time-dependent optical potential well and coupled periodically to different external heat or active baths that enable the cycle to run through all four stages \cite{clemensnature,krishnamurthy2016micrometre,pietzonka2019autonomous}. In this manuscript, we use a similar set-up to engineer a Stirling engine, however, with the utilization of upconverting particles as working substances instead of colloids.

 Optical trapping of upconverting particles \cite{rodriguez2016optical,haro2013optical,kumar2021breaking} has gained recent interest as their potential varies from imaging to therapy\cite{jaque2014nanoparticles,wilhelm2017perspectives,fischer2011upconverting,carlos2009lanthanide}. Generation, control and detection of out-of-plane rotational motions\cite{b1,b2,b3,b5} of single UCP along with the translations in optical tweezers \cite{b4,kumar2020pitch} to explore their micro-rheological and mechanical potentials have also been studied. Recently, Kumar et al. reported a new system of an upconverting particle optically trapped at the absorption resonance of 975 nm, generating a temperature gradient and self-propulsion\cite{kumar2020trapped}. It has been shown that in such translations, the overall motion can be determined by the  propulsion speed, the rotational and translational Brownian motion which leads to a super-diffusive mean square displacement \cite{schachoff2015hot,suresh2022towards} due to this ``activity'' of the particle \cite{schachoff2015hot}. The activity vanishes when the same particle is trapped at a non-absorptive wavelength, 1064 nm. Both the 1064 nm and 975 nm laser beams contribute to trap stiffness of the system whereas 975 nm has an additional effect of generating a temperature gradient across the particle causing the ``activity''. We use this very facet of the working principle to develop a microscopic Stirling engine. 
 
 A Stirling engine functionally consists of closed cycles of isochoric and isothermal processes. At the microscale,  the constant-volume part of the isochoric process is taken care of by the optical confinement, which can further be coupled to periodically varying heat reservoirs to execute the heating and cooling of the same particle. Similarly, the expansion and compression protocols are done by varying the trap stiffness in a fixed heat bath -- reminiscent of the isothermal phase. Notably, the pump wavelength of NaYF$_4$:Yb$^{3+}$,Er$^{3+}$ upconverting particles (975 nm) is employed as one of the trapping wavelengths to generate absorption-enhanced optical heating. Thus, the requirement of external heat baths to couple with the trapped particle can be completely abolished. The isothermal parts of the cycle are executed by delivering a constant power of 975 nm laser to the optical trap and varying the 1064 nm power accordingly.

\begin{figure*}[ht!]
	\centering
		\includegraphics[scale=0.15]{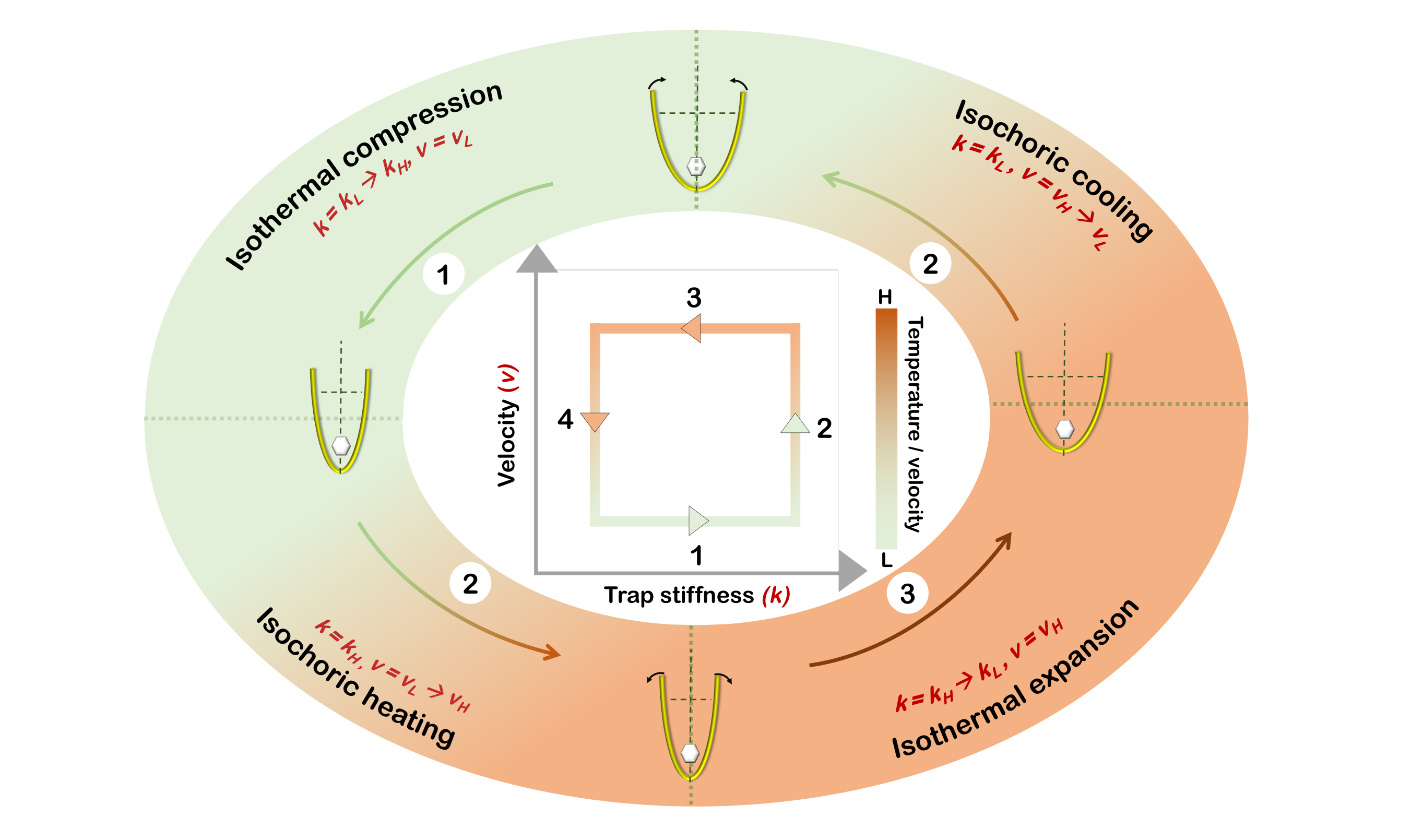}
	\caption{\textbf{Schematic representation of one complete cycle of an engineered microscopic Stirling engine using upconverting particles (see section \ref{sec:results} for details) }}
	\label{mechanism}
\end{figure*}

On the other hand, the isochoric processes are carried out by properly synchronizing both the laser powers so that the total intensity of light and thus the trap stiffness remains constant. Simultaneously, heating and cooling parts are performed by the suitable variation of 975 nm laser power. A schematic representation of one complete cycle of the engine is depicted in Fig. \ref{mechanism}. In what follows, we start by mentioning the basic principles of the engine followed by a detailed description of the experimental setup, analysis, and results.

\section{\label{sec:level1}Theory}
\subsection{An optically trapped upconverting particle subjected to asymmetric temperature profile }

\begin{figure}[h]
	\centering
		\includegraphics[scale=0.2]{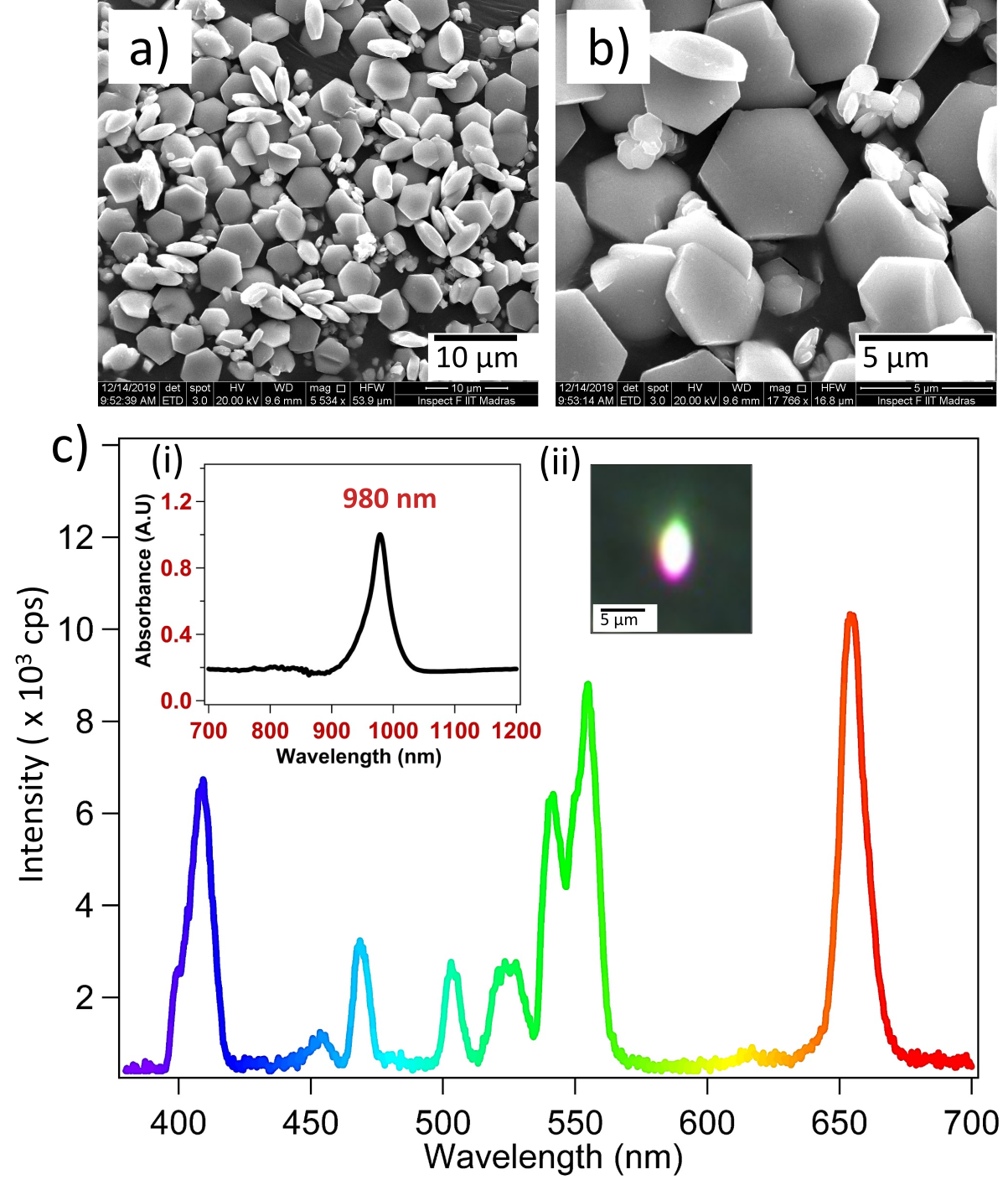}
	\caption{(a-b) The FE-SEM images of UCP are shown to confirm their hexagonal geometry. In (c), the emission spectra of one such UCP trapped with 975 nm laser is shown. Inset shows (i) the normalized NIR absorption spectra of NaYF$_4$:Yb$^{3+}$,Er$^{3+}$ UCPs and (ii) the optically trapped configuration of a single particle with 975 nm laser, where the particle tends to align along its side-on axis.}
	\label{sem}
\end{figure}

\begin{figure}[h]
	\centering
		\includegraphics[scale=0.3]{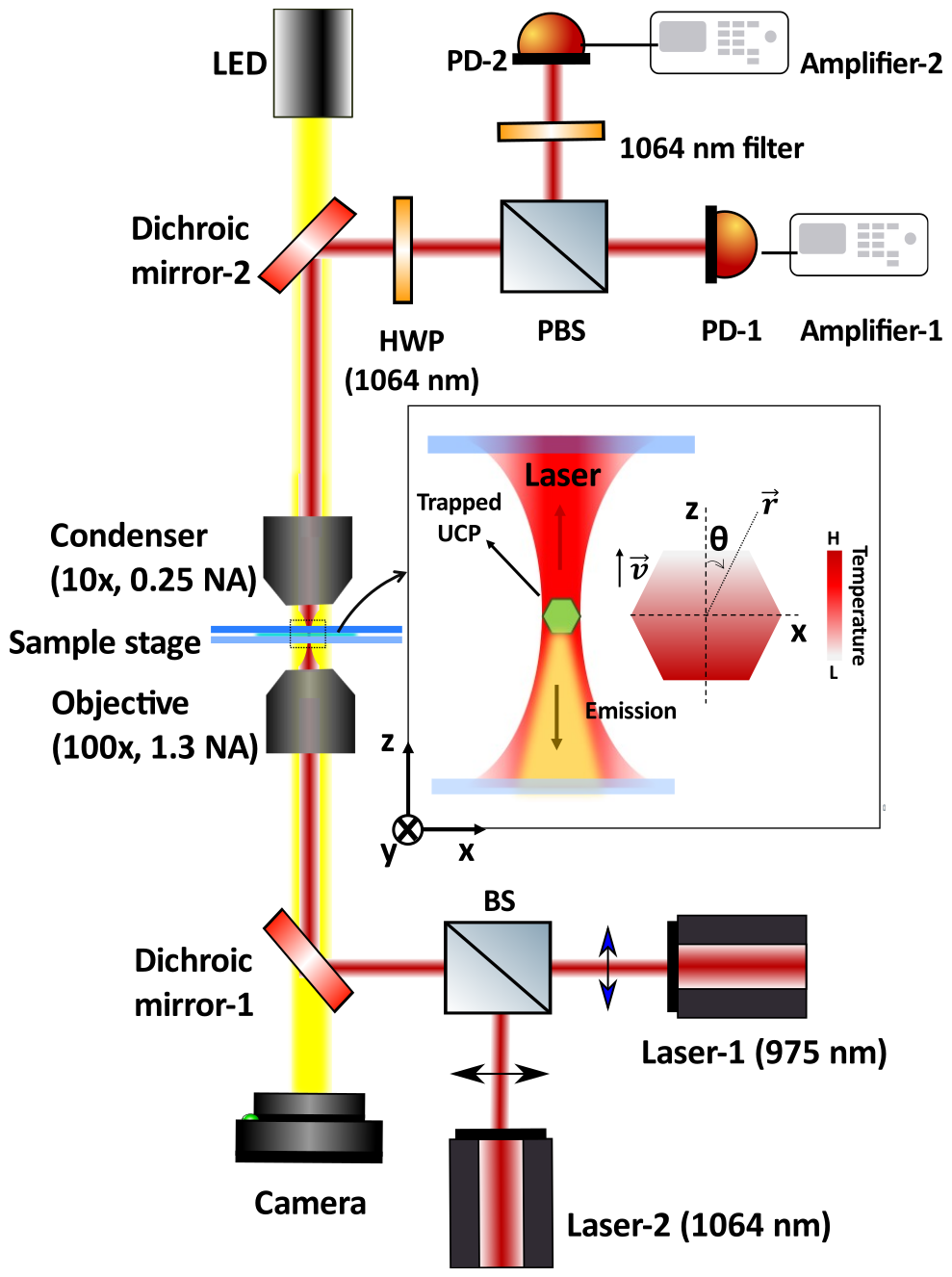}
	\caption{Schematic diagram of the experimental set-up. Both the lasers are tightly focused at the sample plane using 100X objective and the traps are made overlapping to each other. The LED situated at the top illuminates the sample chamber through the dichroic mirror - 2. BS - beam splitter, PBS - polarizing beam splitter, HWP - half-wave plate, PD - photodiode.}
	\label{schem}
\end{figure}

NaYF$_4$:Yb$^{3+}$,Er$^{3+}$ upconverting particles (UCPs) preferentially absorb NIR photons at 975 nm and emit higher energy photons in the visible band with the efficiency of about 2 \% \cite{wen2018advances}. This strong absorption leads to the optical heating of particles by photon-to-phonon conversion in their lattices. When we look at a single UCP (scanning electron microscopy images in Fig. \ref{sem}(a-b)), confined in optical tweezers at the pump wavelength, the size of such particle is much larger than the beam waist of the focused laser beam while it gets trapped in a side-on configuration as shown in the inset of Fig. \ref{sem}(c) \cite{kumar2022estimation}. As a collective result of all these effects, we observe a local temperature gradient across the UCP along the +z direction. It is also observed that the particle predominantly emits in a back-scattered direction (-z - direction, see the schematic in Fig. \ref{schem}) and self-propels in the opposite direction due to the temperature gradient \cite{kumar2020trapped}. This absorption-induced motion can be thought to be an ``active" self-thermophoretic motion inside the optical trap with the tendency of the particle to move from hotter to colder region \cite{tp1,nalupurackal2022hydro}. This will cause the particle to have a complex flow profile with a propulsion velocity, $v$ which is given by \cite{wurger2010thermal,tp2}
\begin{align}
    v = -D_T \nabla T.
    \label{eqth}
\end{align}
Here, $D_T$ is given as $D_T$ = S$_T D$, where $D$ is the thermal diffusion coefficient of the particle, the sign of which determines the thermophoretic direction of the particle. The S$_T$ is the Soret coefficient.

\begin{figure}[h]
    \centering
    \includegraphics[scale=0.25]{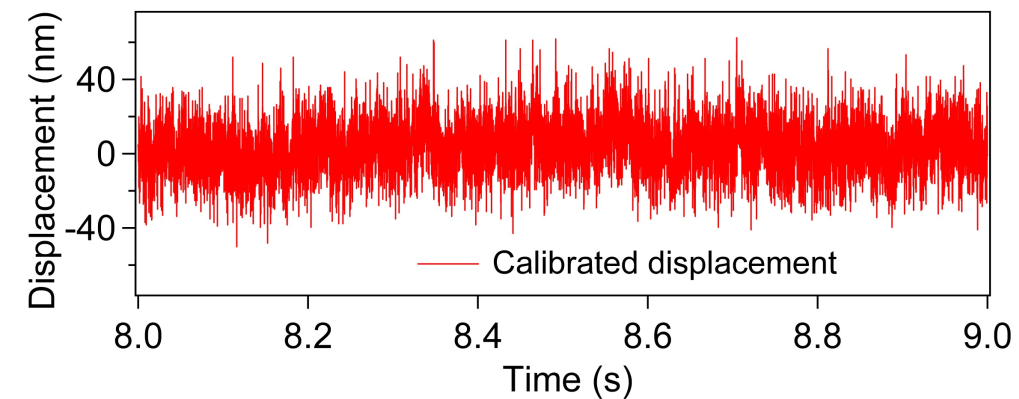}
    \caption{A typical calibrated displacement time series of a UCP along the z-axis, trapped with both 975 nm and 1064 nm laser is shown in the figure. The time series obtained from the isothermal compression phase of the Stirling engine. }
    \label{fig:one_second}
\end{figure}
%-----------------%
%---------%

    The thermophoretic motion generates a velocity $v_0$ in $z$ direction. In our experimental set-up, the particle moves around $40$ nm inside a trap of $1.5$ microns as shown in Fig. \ref{fig:one_second}. Hence, one can safely assume that in the small displacement limit the velocity remains constant. Within this spirit, we therefore can make an assumption that in fluctuation-dominated regimes, the propulsion velocity is an active velocity along the axis of thermophoresis, and thus its information can be extracted from the mean square displacement (MSD) of an ``active'' particle \cite{howse2007self,bechinger2016active,kumar2020pitch,schachoff2015hot,jiang2010active}. For an `active' particle moving with velocity $v_0$ in a random direction in the x-z plane (later we assume that  $v_0$ is almost aligned along the $z$-axis, but the generic theory works for any orientation) in a harmonic trap, one can then write the following Langevin equation \cite{bechinger2016active,kumar2020trapped}
\begin{align}
    \frac{dx}{dt}+\frac{kx}{\gamma}&= v_0 \sin{\theta}+\frac{\eta_x}{\gamma} \\
    \frac{dz}{dt}+\frac{kz}{\gamma}&= v_0 \cos{\theta}+\frac{\eta_z}{\gamma} \\
    \frac{d\theta}{dt}&=\sqrt{D_r} \eta_\theta,
\end{align}
where $k$ is the harmonic trap stiffness, $\gamma$ is the drag coefficient, and $D_r$ is the rotational diffusion constant. The thermal noise $\eta_{x,y}$ is Gaussian with the following statistical properties $\langle  \eta_i \rangle=0$ and $\langle \eta_i(t) \eta_j(t') \rangle =2\gamma k_B T \delta_{ij} \delta(t-t')$. As mentioned above, the activity is captured by the orientation $\theta$ in the x-z plane and the self-propulsion velocity $v_0$ and $\eta_\theta$ is a Gaussian white noise with unit strength. From the property of $\eta_\theta$, it is easy to see that the probability density function of $\theta$ is Gaussian i.e., $P(\theta) \propto e^{-\theta^2/4D_rt}$. From this, one can easily compute the spatial moments (see \cite{howse2007self,solon2015pressure,bechinger2016active,kumar2020trapped}) for $x, y$ and consecutively the radial MSD. Skipping details from \cite{howse2007self,bechinger2016active,solon2015pressure,kumar2020trapped}, one arrives at the following expression for radial MSD
\begin{align}
    \langle \Delta r^2(\Delta t)\rangle = \frac{2k_bT}{k} \left(1-e^{-\frac{2k}{\gamma}\Delta t}\right)  +v_{0}^{2} \left(\frac{1-e^{-\frac{2k}{\gamma}\Delta t}}{\frac{2k}{\gamma}}\right)\left( \frac{4D_r}{4D_{r}^{2}-\left(\frac{k}{\gamma}\right)^2} \right)+ v_{0}^{2} \frac{ 1+e^{-2k\Delta t/\gamma} - 2e^{-((k/\gamma) + 2D_r)\Delta t} }{\left(\frac{k}{\gamma}\right)^2 - 2D_{r}^{2}}.
\end{align}
The expression for MSD above can be simplified further in the limit of small trap stiffness values with $D_r=1/2\tau_r$, where $\tau_r$ is the rotational relaxation time of the trapped UCP \cite{schachoff2015hot}. This results in
\begin{equation}
    \langle \Delta r^2 (\Delta t)\rangle = 4D\Delta t + \frac{v_0^2\tau_r^2}{2}\left(\frac{2\Delta t}{\tau_r}+e^{-\frac{2}{\tau_r}\Delta t}-1\right).
    \label{msd_eq}
\end{equation}
For practical purposes, it can be assumed that the particle is oriented in the x-z plane with a small angle ($\theta$) along the z-axis so that $\cos{\theta} \approx 1$ \cite{schachoff2015hot}.  Henceforth, it can be inferred that $v_0$ is the thermophoretic propulsion velocity along the z-direction, which is the axial velocity of the particle. When the laser trap is switched to a wavelength of 1064 nm, the particle loses its active property and behaves like an ordinary Brownian particle in an optical trap, thus only experiencing the linear diffusive term of Eq. \ref{msd_eq}.
  
\begin{figure}[h!]
	\centering
		\includegraphics[scale=0.2]{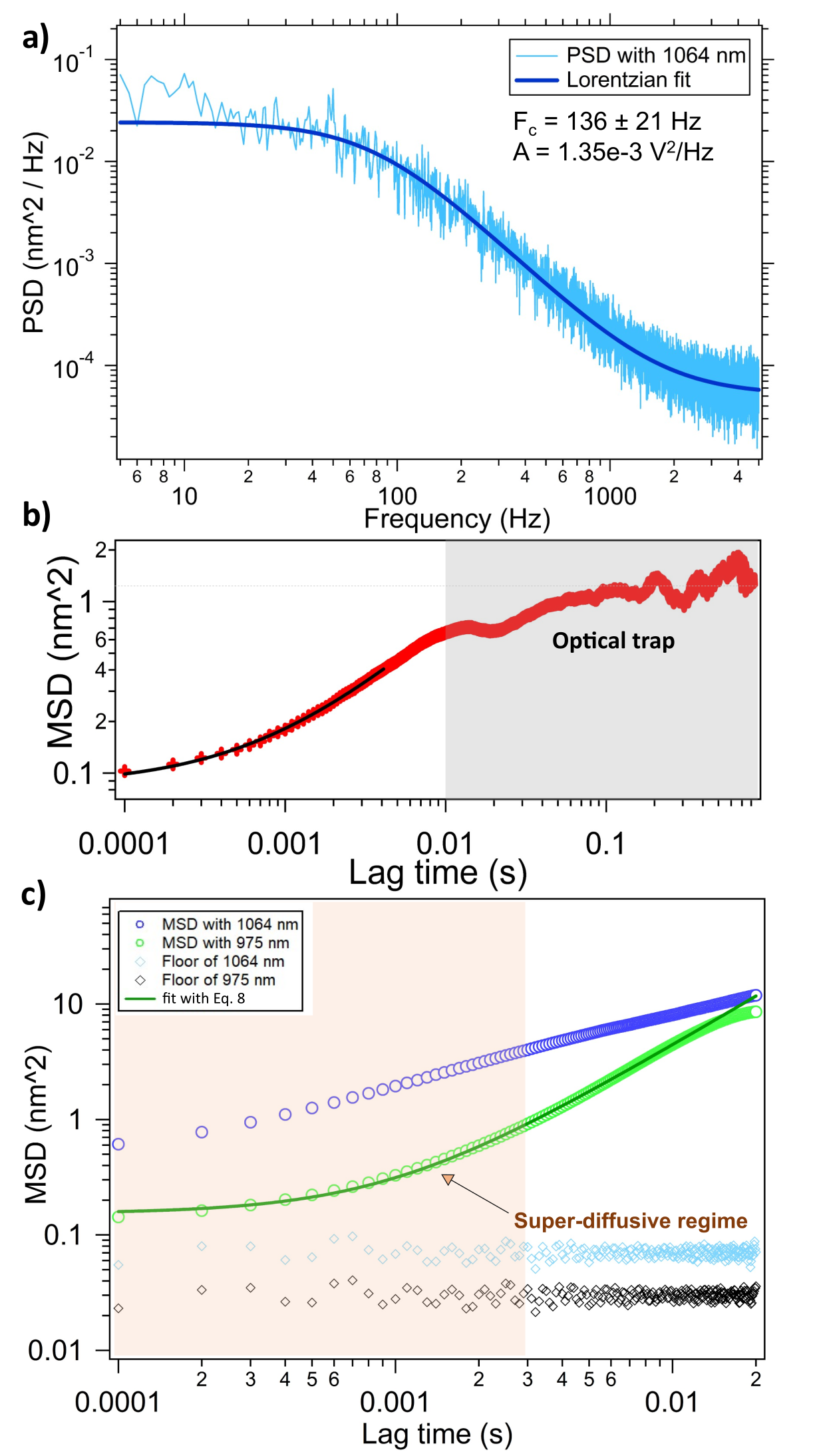}
	\caption{ (a) The calibrated power spectral density of perpendicular displacements (along the z-axis) of the trapped particle with 1064 nm laser is plotted. The displacement sensitivity is found to be $\beta$ = 3.25 $\pm$ 0.13 \textmu m/V. (b) shows the MSD of  the same particle trapped with 975 nm laser. It is fitted to Eq. \ref{msd_eq} and reaches a plateau at large time scales due to optical confinement. In (b), the mean square displacement of the particle along the z-axis, in the x-z plane as a function of time, is plotted and overlaid with the noise floors of both the lasers at the same power. The particle shows pure diffusive behaviour with 1064 nm laser, whereas the same plot shows a super-diffusivity when it is trapped with 975 nm.%
 The curve is fitted with Eq. \ref{msd_eq} to find the axial velocity and relaxation time, gives $v$ = 36.1 $\pm$ 6.3 nm/s and $\tau_r$ = 1.21 $\pm$ 0.32 s.}
	\label{psdmsd}
\end{figure}

  Further, the MSD and the velocity are calibrated with the optical trap by assuming it to be a harmonic trap. The fundamental parameters such as trap stiffness, diffusion coefficient and displacement sensitivity can be extracted from the power spectral density (PSD) of the trapped particle (see Fig. \ref{psdmsd}(a)). The MSD v/s lag-time plot for the particle is shown in Fig. \ref{psdmsd}(b) at active (975 nm) and non-active (1064 nm) wavelengths. Evidently, the active velocity which is extracted from the MSD curve gives the information about the local temperature  $(T)$ profile around the particle (from Eq. \ref{eqth}).

%----------%

%The displacement information of trapped UCP and the corresponding trap parameters can be extracted by power spectral density (PSD) analysis. 
A typical power spectral density extracted from the time series of an optically trapped UCP fitted to a Lorentzian is shown in Fig. \ref{psdmsd}(a). Generally, the PSD of Brownian motion along any axis of an optically trapped particle is given by, 

\begin{align}
    PSD = \frac{A}{(2\pi f)^2 + B},
    \label{psd_eq}
\end{align}
Where $A$ is the amplitude in detector units and $B$ is the square of corner frequency ($f_c$). The trap stiffness ($k$) can be written as \cite{schaffer2007surface},
\begin{align}
    k = 2\pi \gamma \sqrt{B}.
\end{align}
The calibration factor, $\beta$ for the trap, which generally has a unit of $nm\,/\,V$ is given by\cite{schaffer2007surface},
\begin{align}
    \beta = \sqrt{\frac{k_b T}{\gamma A}},
\end{align}
where $k_b$ is the Boltzmann constant and $\gamma$ is the drag coefficient of the particle.  The hexagonal disc-shaped UCP has a diameter of 5 \textmu m and thickness of 1 \textmu m - thus can be approximated geometrically as an oblate spheroid \cite{kumar2020trapped}. Consequently, the drag coefficient becomes $\gamma = 2/3\, \gamma_0$, where $\gamma_0$ is the drag coefficient of a sphere with the same radius as the semimajor axis as the spheroid. The UCPs are placed away from any other surfaces in water.

\subsection{Stochastic work done in a complete Stirling cycle}
The first principles of stochastic thermodynamics are useful to infer the thermodynamic quantities from the microscopic dynamics e.g., work, heat, or entropy production will be fluctuating along any trajectory, say ${x}(t)$ so that one needs to average over many realizations of the process to make a statistical estimation. In this case, work is done since we modulate the potential $U(x,k)$ by changing the stiffness $k(t)$. Following \cite{sekimoto1998langevin,jarzynski1997nonequilibrium,krishnamurthy2016micrometre}, we can then write the microscopic thermodynamical work done in $dt$ time interval as $dW(t)=~\frac{\partial U(\mathcal{C},k)}{\partial k}dk(t)$. For an optically trapped particle, these expressions can be scaled using the trap stiffness of the system. Here, $\mathcal{C}$ is the generalised coordinate (could be position, velocity etc.) of the system. And upon integration, the time-dependent work in the system for a finite time $\mathcal{T}$ is given by \cite{clemensnature,sekimoto1998langevin}
\begin{align}
    W_{T}=\int_0^\mathcal{T}~dt~\frac{\partial U(\mathcal{C},k)}{\partial k}\dot{k}(t).
    \label{eqn1}
\end{align}
It should be noted that this work is due to the time modulation of the trap 
even in the presence of activity \cite{krishnamurthy2016micrometre,dabelow2019irreversibility}
and should be distinguished from the work done due to self-propulsion \cite{shankar2018hidden}.
Work statistics encode important features of an out-of-equilibrium thermodynamic process where tremendous progress has been achieved to measure work fluctuations in microscopic soft and biological systems \cite{seifert2012stochastic,ritort2008nonequilibrium,wen2007force,ciliberto2017experiments,jarzynski2011equalities,seifert2008stochastic}.

 Here, the intensity of both lasers is changed in a linear fashion; over the course of 10 seconds for each process, to ensure the slow variation in dynamics of the particle within the optical trap. Since we are making our observations at much larger time scales than the microscopic timescales corresponding to a particle in an over-damped medium, the trap stiffness will also be directly proportional to the temporal change in the intensity gradient at the time of observation, hence varying linearly over time. Due to the sign conventions, the work is positive (negative) when the energy is supplied (extracted) to (from) the trapped particle system.

Since the upconverting particles have a preferential absorption at 975 nm, we observe a significant temperature gradient across the particles, accounting for its axial motion \cite{kumar2020trapped, schachoff2015hot}. In turn, the temperature of the immediate water layers to the particle also increases and collectively forms a thermal bath. The rise in temperature in the neighbourhood of the particle is accounted for by the optical heating, indicated by its absorption cross-section. It is generally written as \cite{shan2021optical,govorov2006gold},    \begin{align}
   \Delta T = \frac{\sigma I}{4\pi rK},
   \label{eqn2}
\end{align}
where $\sigma$, $I$, $r$, and $K$ respectively are the absorption cross-section, incident laser intensity, the radius of the particle, and thermal conductivity of the surrounding medium. The value of $\sigma$ is determined to be 3.61 $\times$ 10$^{-11}$\,cm$^{2}$ per particle using diffuse reflectance spectroscopy (DRS) for a given number density ($N = 3.55\times10^{15}$\,cm$^{-3}$, see the normalized absorption spectra at the inset of Fig. \ref{sem}(c)). For a power density of 1.60 $\times$ 10$^6$ W/m$^2$ at the beam waist, the maximum temperature rise ($\Delta T$) of the immediate surroundings of the particle is calculated to be approximately 2.75 K.

\section{Experimental system\label{ExperimentalSystem}}
\subsection{Optical tweezers setup}
 The experiments are performed using an optical tweezers setup (OTKB/M, Thorlabs) configured as an inverted microscope. The objective lens of the setup is a 100x, 1.3 NA, oil immersion type, obtained from Olympus. While the 10X condenser lens is an air immersion type from Nikon with 0.25 NA. Two diode lasers of wavelengths 975 nm and 1064 nm, with the same beam-polarization, are employed to create the optical traps at the sample chamber by tightly focusing them with the objective. The laser, namely laser-1 (975 nm) is a butterfly-type laser from Thorlabs, USA with a maximum output power of 300 mW, while laser-2 (1064 nm) is from Lasever, China, and has a maximum output power of 1.5 W. The two optical traps are overlapped with each other. The forward scattered light from the particle and the unscattered light is directed towards the detection systems using the condenser lens and top dichroic mirror as shown in Fig. \ref{schem}. The same dichroic mirror allows the transmission of white light from the top LED to illuminate the sample chamber. This white light is collected by a CMOS camera (Thorlabs) through the bottom dichroic mirror. The combination of a half-wave plate (1064 nm), a polarizing beam splitter (PBS), a band-pass filter (1064 $\pm$ 10 nm), and a pair of Si-photodiodes (PD-1 and PD-2 (DET100A2, Thorlabs), located at the orthogonal ports of PBS) constitute the detection system. Here, most of the forward scattered light of 975 nm laser is made incident on PD-1, while that of 1064 nm laser is directed to PD-2 through a bandpass filter. 
 
 \subsection{Generation of a complete Stirling cycle}
 
 The complete cycle of the microscopic Stirling engine is generated by the careful manoeuvring of intensities of laser-1 and laser-2 over a finite time, $\mathcal{T}$. Both lasers are voltage-controlled current devices and their output powers can be driven by external voltages. We extend the input of the lasers to a NI-DAQ-based voltage source (NI USB-6009, National Instruments, USA). The output power of both lasers is calibrated against the driving voltage from the DAQ using LabView, to select the range of operation. Further, the lasers are simultaneously programmed to generate all four processes of the Stirling engine (see Fig. \ref{timeseries}(a)). The duration of a complete cycle is set to be 40 seconds, divided into 40 independent bins of 1-second duration for the analysis. %A schematic representation of one such complete cycle of operation is depicted in Fig. \ref{mechanism}.
 The individual processes are described in section \ref{sec:results}.
%---------%

\subsection{Preparation of upconverting particles}
Upconverting particles (NaYF$_4$:Yb$^{3+}$,Er$^{3+}$ UCPs) are synthesized through modified hydro-thermal method \cite{kumar2020trapped,kumar2020pitch}. Initially, 1.26 g of yttrium nitrate (Y(NO$_3$)$_3$) and 1.23 g of sodium citrate (Na$_3$C$_6$H$_5$O$_7$) are dissolved in 14 mL of D.I. water and stirred for 15 minutes. At the same time, 0.38 g
of ytterbium nitrate (Yb(NO$_3$)$_3$) and 0.037 g of erbium nitrate (Er(NO$_3$)$_3$) are added and stirred in 21 mL of D.I. water in a separate beaker. The two solutions are mixed and stirred well until a pale white solution is formed. The addition of 1.411 g of sodium fluoride (NaF) in 67 mL D.I. water to the above mixture results in a clear solution, which is then transferred into an autoclave reactor and heated at 200$^\circ$C for 12 hours. The white-coloured powders are formed after the reaction and then collected after washing with ethanol/water, followed by drying at 100$^{\circ}$ C for about 12 hours. The hexagonal geometry of the particles with a diagonal length of 5\textmu m is confirmed from their FE-SEM images as shown in Fig. \ref{sem}(a-b).

The powder containing particles is mixed in D.I water and 20 \textmu L of which is poured on a glass slide (Blue star, English glass, (Blue Star, 75 mm $\times$ 25 mm $\times$ 1.1 mm). This aqueous solution of UCPs is then covered using a coverslip (Blue Star, number 1 size, English glass), which makes the sample chamber. One such UCP is trapped simultaneously with 975 nm laser and 1064 nm laser in the bulk.

\section{Results and discussions}\label{sec:results}
%%---------------------------------------------------
%%---------------------------------------------------
\subsection{Simultaneous determination of axial velocity and trap stiffness}
\begin{figure}[b]
	\centering
		\includegraphics[scale=0.25]{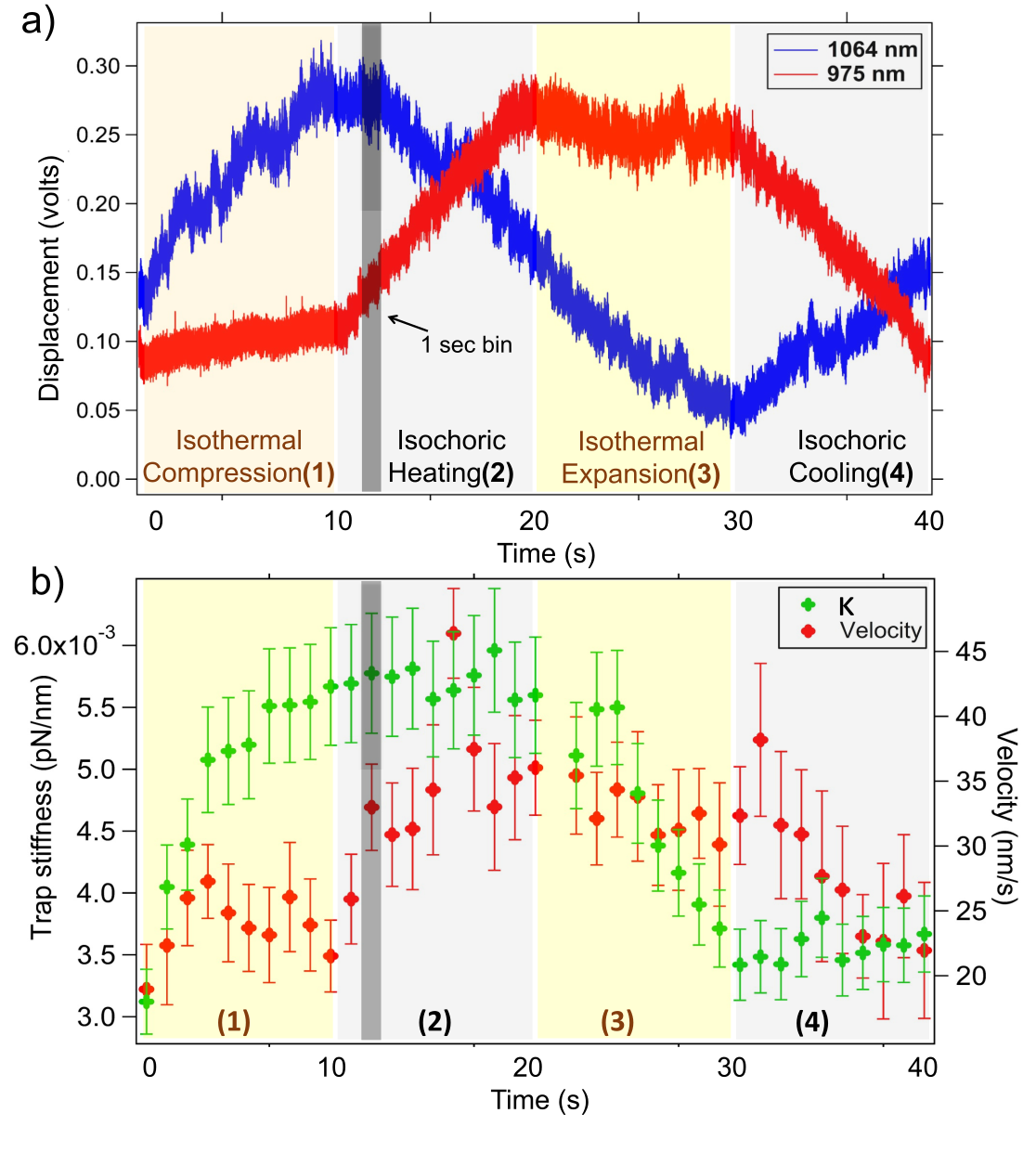}
	\caption{ (a) The displacement time series of the trapped particle along the z-axis, in the x-z plane, obtained using both 1064 nm and 975 nm lasers are shown in the figure. We set the duration of a single cycle to be 40 seconds, divided into ten equal time bins corresponding to each process of the Stirling cycle. In (b), variations of trap stiffness ($k$) and axial velocity of the trapped particle, as obtained from (a), are shown as a function of time over the four stages of a complete cycle.}
	\label{timeseries}
\end{figure}

Fig. \ref{timeseries}(a) shows the axial displacement (z-axis) time series of an optically trapped UCP, obtained simultaneously from the scattered intensities of laser-1 (975 nm) and laser-2 (1064 nm). We use the data corresponding to laser-1 for the purpose of calibration and analysis. 

We design one complete cycle in such a way that each cycle process is 10 seconds long and is further divided into 10 individual bins with a duration of 1 second. Given the data points we collect each second (10 kHz), this requirement is adequate to execute the fast Fourier transform (FFT) and retrieve the trap parameters. The MSD of the particle in each time bin is plotted against the lag-time and fitted to Eq. \ref{msd_eq} to obtain the axial active velocity ($v$) and relaxation time, as shown in Fig. \ref{psdmsd}(b). Similarly, the PSD of the same bin is fitted into a Lorentzian using Eq. \ref{psd_eq} to obtain the trap stiffness. In this manner, the axial velocity of the trapped UCP and trap stiffness is obtained and calibrated throughout one complete cycle and plotted as a function of cycle lag-time in Fig. \ref{timeseries}(b).

It may be realised that both the beams 975 nm and 1064 nm laser are sampling the same particle motion. The same information may be obtained by studying either time series. Hence we use the 975 nm laser time series. Indeed only the calibration factors of the time series of 975 nm laser and the 1064 nm laser would be different for the two cases. The trap stiffnesses have been computed by ascertaining the power spectral density of each 1-second segment of the 975 nm time series, which has been fitted with routine Lorentzian fits shown in Fig. \ref{psdmsd}(b). The method to determine trap stiffness k and the calibration factor has been given in \cite{schaffer2007surface}.

%---------%

We find that our mean square displacement (MSD) of the z-axis motion fits well with the description given in Eq. (\ref{msd_eq}). Indeed it averages out over the 2 rotational angles. Still, one of the out-of-plane (we call pitch angle in the nomenclature of the airlines) angles is confined by the polarization direction in the optical tweezers themselves. As seen in \cite{rodriguez2016optical}, the hexagonal-shaped particle automatically aligns side-on along the direction of linear polarization of the optical tweezers beam. Given that the orientation is locked by the tweezer's beam, a planar description of the system works well. We also assume that the particle moves very little while executing active motion, due to which the extent of anisotropic heating remains constant. 

\subsection{Analysis of a complete Stirling cycle}
The velocity of the trapped UCP and corresponding trap stiffness is plotted as a function of cycle lag-time in Fig. \ref{timeseries}(b). The complete cycle is designed so that each process (leg of a cycle) is 10 seconds long. The stochastic work done in five such cycles is estimated from the displacement time series of the particle obtained from the scattered intensities of 975 nm laser and plotted in Fig. \ref{cycle}(a). Here, we run the cycles in an anti-clockwise direction as shown in Fig. \ref{cycle}(b) and determine $k$ and $v$ values for one such cycle. The calibrated values of active velocity and trap stiffness yield the $k-v$ diagram of the engine over a full period (see Fig. \ref{cycle}(b)). This plot is analogous to the V-T diagram of a macroscopic Stirling engine. 
\subsubsection{Isothermal compression}
We start the cycle by exploring the isothermal compression of its working substance (here, the confinement volume of trapped UCP). UCP is trapped with both lasers at the same power in the initial configuration. Then, the power of 975 nm laser is kept constant and that of 1064 nm is increased gradually over time, so as to follow a linear trend. This process over the first ten seconds of the cycle ensures that the optical heating of the trapped particle is retained constant and the optical trap stiffness is increased linearly (from $k_L$ to $k_H$). The observations are plotted for the first ten seconds in Fig. \ref{timeseries}(b). During this process, the temperature of the bath is constant ($T_L$) and the corresponding active velocity of the particle  will be constant ($v_L$)  according to Eq. \ref{eqth}. This can be attributed to the constant activity delivered by the constant laser power of 975 nm laser. On the other hand, the trap stiffness is increased in a linear fashion with the increase of 1064 nm laser power, providing the system with a less final volume. We ensure that the trap stiffness follows a linear modulation in time. 

\subsubsection{Isochoric heating}
Since isochoric heating of UCP demands a constant trap stiffness ($k_H$) and a simultaneous increase in the heat bath's temperature (from $T_L$ to $T_H$), synchronized variation of both the laser powers is required. The heating part is done by increasing 975 nm laser power, the additional contribution of the same on trap stiffness is offset by decreasing 1064 nm laser power. This process results in an increase in active velocity ($v_L$ to $v_H$) keeping the trap stiffness constant to make it isochoric. The trap stiffness value is constant in time throughout this process which indicates $\dot{k} =0$ (resulting in null work done). The results are depicted from 11 to 20 seconds in Fig. \ref{timeseries}(a) and (b). 

\subsubsection{Isothermal expansion}
Isothermal expansion occurs once the temperature of the heat bath is sufficiently high in a Stirling engine ($T_H$). In this process, the power of 975 nm laser is kept constant to make it isothermal and that of 1064 nm laser is decreased linearly in time to observe the decrease in trap stiffness of the system. The active velocity of trapped UCP remains constant ($v_H$) and the trap stiffness decreases to a minimum value ($k_L$) at the end of the process (21 to 30 seconds of Fig. \ref{timeseries}(b)).

\subsubsection{Isochoric cooling}
The cycle is concluded by making the temperature of the bath its initial value, $T_L$. Here, again the power of both 975 nm and 1064 nm lasers are varied simultaneously but in an opposite fashion to that of isochoric heating. The 975 nm laser power is decreased linearly to bring down the temperature and active velocity of the particle to their initial values ($T_L$ and $v_L$). And the corresponding reduction in trap stiffness is compensated by increasing 1064 nm laser power simultaneously, as depicted in the last 10 seconds of Figs. \ref{timeseries} (a) and (b).   

\subsection{Calibration of stochastic work done in multiple cycles}
The work done is calculated using Eq. \ref{eqn1} and calibrated with the optical trap. The trap stiffness $k$ of the system is set to vary linearly with time over a period $0<t<\mathcal{T}$, the rate of which can be extracted from the slope of isothermal processes (see Fig. \ref{timeseries}(b)). Upon substituting the proper parameters for calibration, the thermodynamical work in Eq. (\ref{eqn1}) can be written as
\begin{align}
    W_\mathcal{T}=\frac{k_b T}{2\gamma A}\int_0^\mathcal{T}~\dot{k}(t)\,z^2(t)~dt,
    \label{eqn3}
\end{align}
where we have assumed $U=\frac{1}{2}k(t)z^2$ with $k(t)$ as a linear function in time, which implies $\dot{k}(t) =$ C, a constant. We also get from Fig. \ref{timeseries}(b), C$_{compression} = 0.36 \pm 0.05\,\, fN/nm\cdot s$, C$_{expansion} = 0.43 \pm 0.03\,\,fN/nm\cdot s $. From Eqs. (\ref{eqn1}) and (\ref{eqn3}), it is obvious that the isochoric transitions do not contribute to the work done. The work done is conventionally taken as positive during isothermal compression as the energy is supplied to the system, thereby signing the work done during expansion as positive.

%---------------------------------------------------------%
\begin{figure}
	\centering
		\includegraphics[scale=0.25]{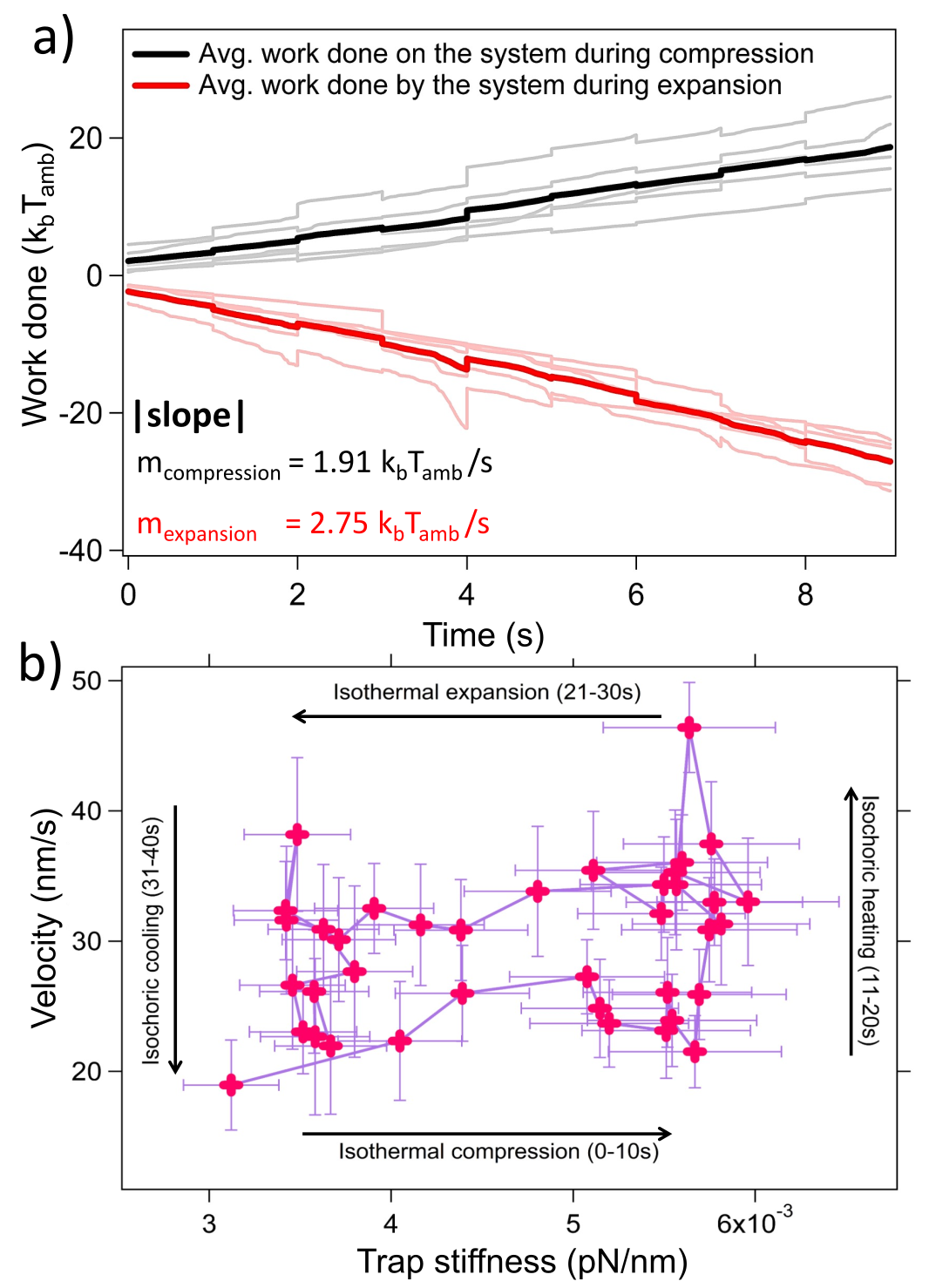}
	\caption{a) The average work done over five cycles of isothermal processes of the engine is shown as bold red and black curves -- corresponding to the compression and expansion stages. The individual stochastic work done over all five cycles is separately overlaid on the same plot. Each cycle is taken independently and referred to as zero work at the beginning of isothermal compression and expansion to compare the magnitudes. The bold black and red curves are the average work done over all the cycles. The linear evolution of the work done is quantified by its slope (m). For the process of isothermal compression, the aforementioned slope is calculated to be $m_\text{compression} = 1.91 k_bT/s$, whereas in the expansion it becomes,  $m_\text{expansion} = 2.75 k_bT/s$, where  T$_{amb}$ = 298 K. The difference in the magnitude of the slopes indicates the measure of useful work which could be extracted from the heat engine.  (b) This figure shows the $k-v$ diagram of a Stirling cycle (shown in Fig. \ref{cycle}). This is equivalent to the $T-V$ diagram of a macroscopic engine. 
	}
	\label{cycle}
\end{figure}
%--------------------------------------------%
 
 We determine the work done during compression and expansion for five individual cycles as shown in Fig. \ref{cycle}(a). It can be seen that the work done is stochastic, as the system is subjected to fluctuations. The work done along each realization and their average is shown in the same figure. Each bin of the displacement time series is calibrated and  the square of which is integrated altogether with respect to time to obtain the stochastic work done. The obtained work is then normalised with the ambient temperature, T$_{amb}$ = 298 K. The work done during isothermal expansion and compression evolve linearly with the process time. The averaged stochastic work is then characterized by the slope of these two lines, given by $\lvert m_\text{compression}\rvert = 1.91\, k_bT/s$ and
$\lvert m_\text{expansion}\rvert = 2.75 \,k_bT/s$. Here, by sign convention, we take work done on the system (by the system) to be positive (negative). Hence, the useful stochastic work that can be extracted from the engine is indicated by the difference in the magnitudes of the slopes. %($\Delta m$).% It can be seen that $\Delta m = (W_\text{expansion}-W_\text{compression})/ \tau $, where $\tau$ is the duration of compression or expansion. 

The optical tweezers is a non-conservative system, primarily because the radiation pressure induces circulatory probability currents inside the trapping potential \cite{roichman2008optical}. However, this circulatory velocity is proportional to the intensity of the light applied. We find in the condition that when the total intensity is held constant (isochoric process) but the 975 nm laser intensity is changed, the drift velocity changes as shown in Fig. \ref{timeseries}(b). This change in velocity cannot be explained by a constant circulatory probability flow. Thus we believe that non-conservative effects due to the optical tweezers is not causing the drift velocity. 

Moreover, since the z displacement of the particle itself is directly influenced by the heating, and only the active particle needs to be heated in this case, the efficiency of the engine should be higher than trying to heat up the entire surrounding bath. However, the exact calculation of the efficiency of the system is beyond the scope of the present manuscript. 

\section{Conclusions and perspectives}\label{sec:conclusion}

To summarize and conclude, we have attempted to engineer a microscopic, closed-cycle Stirling engine with upconverting particles. The absorption and optical heating properties of UCPs at 975 nm eliminate the requirement of external heat baths for the engine to function. We introduce active velocity of the particle ($v$) and trap stiffness ($k$) as the indicative parameters of the engine and show that they follow a theoretically anticipated trend over a complete cycle. We also demonstrate that the active velocity is a measure of the temperature of the particle's absorption-generated heat bath and hence we report the $k-v$ diagram of the engine, which is analogous to the $V-T$ diagram of a macroscopic Stirling cycle.  Finally, we determine the stochastic work done over five such cycles by taking the notion of a single particle that is subjected to a time-dependent optical trap and periodically coupled to different heat baths.

The optical heating of a confined UCP can be controlled precisely and timely without causing any plasmonic effects or other higher-order effects, which constitutes one of the major advantages of using an absorbing particle as the working substance. This external control over temperature ensures that the microscopic, thermodynamic variables can be parameterized more accurately in experiments. We thus believe that our work can surmount one of the major challenges in colloidal heat engines, which is the temperature control in isochoric processes of information heat engines. Furthermore, the particles we use can also be employed to study the non-steady heat transfer between different potential wells in optical tweezers systems under laser irradiation. This renders a singular appeal of our experimental system with upconverting particles to study various  thermodynamical properties in the out-of-equilibrium conditions -- a subject that has become a topical interest in statistical physics \cite{ciliberto2017experiments}.
 The extent of inertia at the micro hydrodynamical systems is very less but it is actually applicable to all systems at these length scales. We can however envisage systems where the active particle is made to push against an elastic membrane which then does work.

\emph{Acknowledgements.---}
We thank the Indian Institute of Technology, Madras, India for their seed and initiation grants. This work was also supported by the DBT/Wellcome Trust India Alliance Fellowship IA/I/20/1/504900 awarded to Basudev Roy. Arnab Pal gratefully acknowledges research support from the DST-SERB Start-up Research Grant Number SRG/2022/000080 and the Department of Atomic Energy, India.

\bibliography{ref}

\end{document}